\documentstyle[prl,aps,preprint]{revtex}
\begin{document}

\newcommand{\bee}{\begin{equation}}
\newcommand{\ene}{\end{equation}}
\newcommand{\nn}{\nonumber}
\newcommand{\ra}{\rightarrow}
\newcommand{\bea}{\begin{eqnarray}}
\newcommand{\ena}{\end{eqnarray}}
\newcommand{\bear}{\begin{array}}
\newcommand{\enar}{\end{array}}
\newcommand{\lb}{\label}
\newcommand{\fr}{\frac}
\newcommand{\pa}{\partial}
\newcommand{\na}{\nabla}
\newcommand{\p}{\! +\!}
\newcommand{\m}{\! -\!}
\newcommand{\bef}{\begin{figure}}
\newcommand{\enf}{\end{figure}}

\newcommand{\al}{\alpha}
\newcommand{\be}{\beta}
\newcommand{\g}{\gamma}
\newcommand{\e}{\epsilon}
\newcommand{\ve}{\varepsilon}
\newcommand{\la}{\lambda}
\newcommand{\de}{\delta}
\newcommand{\bde}{{\bf\delta}}
\newcommand{\si}{\sigma}

\newcommand{\z}{\zeta}
\begin{center}
   {\Large \bf  Dynamics of particle deposition on a disordered substrate:
   I. Near-Equilibrium behavior.
}
\end{center}
\begin{center}
Yan-Chr Tsai $\footnote[2]{Present address: Department of Physics and Astronomy,
 University of Pennsylvania,  Philadephia,   PA  19104-6396}$ and Yonathan Shapir
 \vskip 0.25in
Department of Physics and Astronomy,
University of Rochester, Rochester, NY 14627

\end{center}
\begin{abstract}

A growth model which describes the deposition of particles
(or the growth of a rigid crystal) on a disordered substrate is investigated.
The dynamic renormalization group is applied to the stochastic growth equation
using the Martin, Sigga, and Rose formalism. 
The periodic potential and the quenched disorder,
upon averaging,  are combined into
a single term in the generating functional. Changing the temperature
(or the inherent noise of the deposition
process) two different regimes with a transition between them at 
$T_{sr}$,  are found:
for $T>T_{sr}$ this term is irrelevant and the surface has the scaling
properties of a surface 
growing on a flat substrate in the rough phase. The
height-height correlations behave as $C(L,\tau)\sim \ln [L f(\tau/L^2)]$. While
the linear response mobility is finite in this phase it does vanish as 
$(T-T_{sr})^{1.78}$ when $T\ra T_{sr}^+$. For $T<T_{sr}$ there is
a line of fixed-point for the coupling constant.
The surface is super-rough: the equilibrium correlations function behave
as $(\ln L) ^2$ while their short time dependence is $(\ln \tau) ^2$
with a temperature dependent dynamic exponent
$z=2[1+1.78(1-T/T_{sr})].$

While the linear response mobility vanishes on  large  length scales, 
its scale-dependence leads to a non-linear response. For a small
applied force $F$ the average velocity of the surface $v$ behaves as $v \sim 
F^{1+\zeta}$. To first order $\zeta=1.78(1-T/T_{sr})$.  At the transition, 
$v\sim F/(1+C|\ln (F)|)^{1.78}$ and the crossover to the behavior to $
T<T_{sr}$ is analyzed.  These results also apply to two-dimensional vortex
glasses with a parallel magnetic field.
\end{abstract}

\newpage                                                             
\section{Introduction}
\label{sec51}

Much  progress has been achieved recently in the understanding of surface growth in
processes
of deposition, sedimentation, epitaxial growth, solidification, etc \cite{BN,GR1,krug}. 
A few years earlier the static and the dynamics properties of roughening
of crystalline surfaces  were elucidated \cite{CW,NG}. Recent investigations have
concentrated on  the connections between surface roughening due to thermal
fluctuations on one hand and that due to the kinetic growth
itself,  on the other hand.
The underlying discrete structure of the particles (or the lattice)
may lead to a kinetic phase transition  between smooth and rough phases or
between two rough phases with distinct scaling properties \cite{BN}. 
Since the underlying discrete structure is relevant at  low temperature
(or a low noise regime  
in the  deposition)
one cannot escape the question on how disorder in the substrate might modify
the surface properties.
 
The effect of the substrate disorder on the dynamics of the growing
surface is the subject of our analysis \cite{TS}. We address this
issue using the dynamic renormalization  group (RG) applied to
the stochastic growth equations using the   Martin, Sigga, and Rose
(MSR) formalism\cite{MSR}.

In the present paper we address the  dynamics near-equilibrium. This
regime is characterized by a very slow rate of deposition such
that the system is very close to thermodynamic equilibrium.
In this regime  only  slight modifications from equilibrium are
considered.  
In particular the fluctuation-dissipation theorem (FDT) \cite{MM,SMS} and the Einstein
relation ( between the mobility, the diffusion constant and the temperature)
both hold. 

Far-from equilibrium,  the growth equation does not obey the FDT. The symmetry
under $h\ra - h$ ($h$ is the height of the surface) and 
time reversal $t\ra -t$   are  broken. The most relevant additional term,
as shown by Kardar, Parisi, and Zhang (KPZ) \cite{KPZ} is due to the lateral growth
of the oblique surface. 
The behavior far-from equilibrium will be the subject of a following paper.

In  general, the scaling properties of the growing surface are manifested in the height-height 
correlation function:
\bee
C(L,\tau)=<(h(\vec x+ \vec L, t+\tau)-h(\vec x, t))^2 > 
\ene
or the corresponding surface width:

\bee
W(L,\tau) =[C(L,\tau)]^{1/2},
\ene

which  obeys an asymptotic behavior of the form:
\bee
W(L,\tau) = L^{\alpha} f(\tau/ L^z).
\label{eq:nh} 
\ene

In this expression $\alpha$ is the roughness exponent which characterizes
the  extent of the roughness of the surface and $z$ is the dynamic exponent.
$f(x)$ is a scaling function which approaches a constant 
for large $x$.
For small $x$ ( $\tau << L^z)$: $f(x) \sim x^{\beta}$ where $\beta=\alpha /z$.
At early stage of the growth the surface roughness increases as $W(\tau)
\sim \tau ^{\beta}$ while for  $\tau >> L^z $, $W$ depends only on
$L$ and behaves as $L^{\alpha}$.

In the absence  of any disorder in the substrate the near-equilibrium
behavior was analyzed extensively in the context of surface roughening.
The original work  was  due to Chui and Weeks (CW) \cite{CW} and this system was further
analyzed by Nozieres and Gallet (NG) \cite{NG}. Their most important findings were
as follows:  

In the high temperature rough phase 
\begin{equation}
C(L,\tau) \sim \ln [L f(\tau /L^z)]
\label{eq:jj} 
\end{equation} 
which corresponds to $\alpha =0$, $\beta =0$ and $z=2$.  
The scaling form  of the correlation  function in Eq.~(\ref{eq:nh}) can  not  
apply  to  that of  Eq.~(\ref{eq:jj}), since both $\alpha $ and
$\beta$ vanish but come with a finite ratio. 

In this phase
the effect of the discreteness (or the lattice) is not
relevant.   This behavior is equivalent to that of a free-surface in which  the
surface tension is the only interaction determining its properties.

In this  regime the macroscopic mobility defined as the ratio 
between the average velocity $v=<\frac{dh}{dt}>$ and the "force"  
$F$ driving the surface, is finite.

In the smooth phase $C(L)$ is independent of $L$, and the mobility vanishes. The mobility has 
a finite jump from a finite value to zero at the roughening temperature.
The growth process at low temperatures is by nucleation of higher "islands"
on top of the smooth surface. This "activated" growth has drastically 
different dynamic properties which are determined by the diffusion
of the  deposited particles on the surface and their attachment to
the "islands". 

While we study here the surface properties in deposition of cubic
(or tetragonal) rigid particles, our study applies  as well to the growth 
of crystalline surfaces if the rigidity of the solid is large enough. 
Our theory will apply if the surface height is smaller than the scale
on which the random deviations  in the substrate cease to  affect the
positions, along the growth direction,  of the lattice ions. This scale
will be larger the stronger  is the Young's modulus which measures the
longitudinal rigidity.

It turns out that the same stochastic equation of motion also describes
the behavior of other random two dimensional systems. The most important
case is  a system
of vortex lines in a superconducting  film with the applied magnetic field
parallel to the film.
(Charge density waves at finite temperature is  another such a system).
Therefore the conclusions of our investigations also apply to 
2D vortex-glasses \cite{FFH89,FFH891,nat,ST}. 
We shall come back to these implications in the last section.

The outline of this paper is as follows:
In the  section II we present the stochastic 
equation of growth and the related MSR generating
functional. In section III the RG scheme is outlined and the recursion relations are
derived.  Section IV is devoted to the discussion of the results and their
physical implications. The final section V is dedicated  to a summary 
of the important conclusions. In the appendices we provide more details of the RG 
calculations.  A short letter announcing the most important results was published elsewhere \cite{TS}.

\section{ The equation of motion and the associated generating
functional}
\label{sec52}
The prototypical  paradigm for the simplest deposition process  is the Edwards and
Wilkinson (EW)
model \cite{EW} for the sedimentation of  granular particles. 
The  continuum limit form of their  equation of motion  for
the height $ h(\vec x, t) $ is: 
\begin{equation}
\tilde {\mu} ^{-1} \frac{\partial}{\partial t} h(\vec x ,t)=\nu \nabla ^2
h(\vec
x,t)+\tilde{\zeta}(\vec x,t)  + \tilde{F}
\label{eq:ew}
\end {equation}
$\tilde{\mu}$ is the microscopic "mobility" of the upper surface,
$\nu$ is the " diffusion constant" for the particles on 
the surface,  $\tilde{F}$ is  proportional to the averaged
deposition rate which is very small (large deposition rate will be discussed
in a second paper), and 
$\tilde{\zeta}(\vec x,t)$ is the local fluctuation from the averaged  deposition 
rate, which obeys: 
\bee
<\tilde{\zeta}(\vec x,t) \tilde{\zeta}(\vec x^{'} ,t^{'})>=2\tilde{D} \delta
^{2}(\vec x -\vec x^{'})
\delta (t^{'}-t). 
\ene

We can define the  effective " temperature" of this system
by the  Einstein 
relation: $T=\tilde{D}\tilde{\mu}$.

If the discrete nature of the  particles  is taken into account
 the height $h$ of every column of particles must 
be an integer multiple of the vertical size of the particle $b$.
This discrete constraint  leads to a periodic $\delta$-function
potential on $h$. This periodic potential  may be 
expanded in Fourier series of which only the basic harmonic is relevant.
On a flat substrate there will be an additional term of the form
$\gamma \frac{\tilde{y}}{a^2} \sin (\gamma h(\vec x, t))$ on the r.h.s.
of Eq.~\ref{eq:ew} ($\gamma= \frac{2\pi }{b}$
and $ \frac{\tilde{y}}{a^2}$ is the amplitude of the periodic potential).

In the presence of  a random substrate the minima of the potential 
will be randomly, and independently, shifted for each column. 
Hence the equation of motion becomes:

\begin{equation}
\tilde{\mu} ^{-1} \frac{\partial}{\partial t} h(\vec x ,t)=\nu \nabla ^2
h(\vec
x,t)- \gamma \frac{\tilde{y}}{a^2}  \sin[\gamma [h(\vec x,t) + d(\vec x)]]+
 \tilde{ \zeta}(\vec x,t)  + \tilde{F}
\ene
 $d(\vec x) $ is  the local deviation of the disordered substrate as depicted 
in Fig. 1. The associated "phase" $\Theta (\vec x)= \frac{2\pi d(\vec x
)}{b}$ is  uniformly distributed between $0$ and $2\pi$,
and is  uncorrelated for different location $\vec x$ on the substrate.

Then the equation of the growth process becomes:
\begin{equation}
\tilde {\mu} ^{-1}\frac {\partial h(\vec {x},t)}{\partial
t}=\tilde{F}+\nu(\nabla ^2
h(\vec {x},t))+\frac {\gamma \tilde{y}}{a^2}\sin
[\gamma h(\vec {x},t)+\Theta (\vec {x})] +\tilde {\zeta} (\vec
{x},t),	       
\label{eq:motion}
\end{equation}      
where  $a$ is the lattice constant 
in the horizontal plane.
    
To investigate this stochastic equation systemically,
one can utilize the MSR (Martin, Sigga, and Rose) formalism 
\cite{MSR} by introducing an auxiliary field $\tilde{h}$ to force
Eq.~(\ref{eq:motion})  through a functional integral representation
of a $\delta$-function. 	
 The generating
functional  for Eq.~(\ref{eq:motion}) takes the form as 
(after averaging over $\zeta
(\vec x, t)$):                  
\begin{equation}
Z_{\Theta }[\tilde {J},J]=\int {\cal D}\tilde {h }{\cal D}h
\exp \{ S_{0}[\tilde {h },h]
+S_{I}[\tilde {h },h ]+\int d^2 x dt (\tilde {J}\tilde
{h}+J h )\},
\label{eq:gen}
\end{equation}
where
\begin{equation}
S_{0}[\tilde {h },h ]=\int d^2x dt [\tilde {D}\tilde {\mu} ^2
\tilde {h
}^2-\tilde {h }(\frac {\partial }
{\partial t}h -\tilde {\mu}\nu \nabla ^2 h )],
\label{eq:s0}
\end{equation}
\begin{equation}
S_{I}=\int d^2 x dt [\frac {\tilde {\mu} \gamma \tilde{y}}{a^{2}} \tilde
{h}\sin(\gamma h (\vec {x},t)+\Theta (\vec {x}))].
\label{eq:gen2}                               
\end{equation} 
The generating functional 
$Z[J,\tilde {J}]$ can be directly averaged over the quenched disordered  $d(\vec x)$
because $Z[J=\tilde{J}=0]=1$ \cite{janssen}. One may calculate any  averaged correlation
and response  function
by differentiating the generating functional with respect to the
current $J$ or auxiliary current $\tilde {J}$  and setting $J=\tilde{J}=0$.

After averaging over the disorder  the effective
generating functional reads:
\begin{eqnarray}
\langle Z_{\Theta }[\tilde {J},J] \rangle _{disorder}&=&\int
{\cal D}\tilde {h }{\cal D}h  \exp\{
\int d^2 x dt [\tilde {D}\tilde {\mu} ^2 \tilde {h }^2 -\tilde {h
}(\frac{\partial }{\partial  t}h -\tilde {\mu} \nu 
\nabla ^2 h )] +\nonumber \\
&&\frac {\tilde {\mu} ^2 \gamma ^2 \tilde {g}}{2a^2}\int \int
d^2x dt dt' \tilde
{h}(\vec {x},t) 
\tilde {h }(\vec {x},t')  \cos (\gamma (h (\vec {x},t)-h
(\vec {x},t'))\},\nonumber  \\
&&
\label{eq:dgen1}
\end{eqnarray}                       
where $\tilde{g}=\tilde{y}^2$.
If we choose $\tilde{D}= T\tilde{\mu} ^{-1}$,  the system will evolve
into the configurations weighted by a  Boltzmann factor
$e^{-H/ T}$, which obeys the fluctuation dissipation theorem 
(FDT) \cite{MM}.
To simplify the calculation, we redefine those physical
parameters as:     
$\tilde {\mu} \nu = \mu, \tilde {D}\nu ^{-2}=D, \tilde {g}=g \nu
^2, \tilde {\bar {\nu}}=\bar {\nu} \nu ^2$, $F=\frac
{\tilde{F}}{\nu}$.
Then the  equation of motion becomes:
\begin{equation}
\mu ^{-1}\frac {\partial h(\vec {x},t)}{\partial t}=F+(\nabla ^2
h(\vec {x},t))+\frac {\gamma y}{a^2}\sin 
(\gamma h(\vec {x},t)+\Theta (\vec {x})) +\zeta (\vec {x},t).
\label{eq:moti}
\end{equation}
Here $\langle \zeta (\vec {x_{1}},t_{1}) \zeta (\vec
{x_{2}},t_{2})\rangle =2 {D} \delta
^{(2)}(\vec {x_{1}}-\vec {x_{2}}) \delta (t_{1}-t_{2}).$               
The resulting  generating
functional  for the present case reads:                 
\begin{equation}
Z_{\Theta }[\tilde {J},J]=\int {\cal D}\tilde {\phi }{\cal D}\phi
\exp \{ S_{0}[\tilde {\phi },\phi]
+S_{I}[\tilde {\phi },\phi ]+\int d^2 x dt (\tilde {J}\tilde
{\phi
}+J \phi )\},
\label{eq:ger}
\end{equation}
where,
\begin{equation}
S_{0}[\tilde {\phi },\phi ]=\int d^2x dt [D\mu ^2 \tilde {\phi
}^2
-\tilde {\phi }(\frac {\partial }
{\partial t}\phi -\mu \nabla ^2 \phi )],
\label{eq:s02}
\end{equation}
\begin{equation}
S_{I}=\int d^2 x dt [\frac {\mu \gamma y}{a^{2}} \tilde {\phi
}\sin
(\gamma \phi (\vec {x},t)+\Theta 
(\vec {x})],
\label{eq:ger2}
\end{equation}

where $h(\vec x,t) =\phi(\vec x, t)$  and $\tilde {h}(\vec x,t)  
\sim \tilde {\phi}(\vec x,t)$. 
In the same way, we arrive at  the averaged effective generating
functional:
\begin{eqnarray}
\langle Z_{\Theta}[\tilde {J},J] \rangle _{disorder}&=&\int
{\cal
D}\tilde {\phi }{\cal D}\phi  \exp\{
\int d^2 x dt [D\mu ^2 \tilde {\phi }^2 -\tilde {\phi }(\frac
{\partial }{\partial  t}\phi -\mu 
\nabla ^2 \phi )] +\nonumber \\
 &&\frac {\mu ^2 \gamma ^2 g}{2a^2}\int \int d^2x dt dt' \tilde
{\phi}(\vec {x},t) 
\tilde {\phi }(\vec {x},t')  \cos (\gamma (\phi (\vec {x},t)-\phi
(\vec {x},t'))\}.\nonumber  \\
&&
\label{eq:dgen}
\end{eqnarray}

In the generating functional in Eq.~(\ref{eq:dgen}), the term which contains the $\cos [\gamma
(\phi (\vec x,t)-\phi (\vec x, t^{'})]$ is non-local in
time and will be responsible for the creation of   non-trivial
Edwards-Anderson correlations.

In the renormalization process it turns out that this term generates
a new " quadratic" term, also non-local in time, of the form:

\begin{eqnarray}
\frac {1}{2}\mu ^2 \bar {\nu} \int dx dt dt ^{'}
\nabla \tilde {\phi}(x,t) \nabla \tilde {\phi}(x,t')
\end{eqnarray}

We therefore add this term to the generating functional and will
follow the flow of $\bar {\nu} $ as well (as shown below it will play a crucial role
in altering the long-range height-height correlations.)
                                      
Here  we  treat the
the last  term in Eq.~(\ref{eq:dgen}) as  a  perturbation
to the free action, and expand the theory in orders of $g$ and
$\delta
(=\frac {\gamma^2 D \mu}{4\pi }-1)$.
The RG scheme will be discussed in the next 
section and details are given in the appendices.

\section{The Renormalization Scheme and the Recursion Relations}
\label{sec53} 
The renormalization group scheme we follow is based on the sine-Gordon
field theory developed by Amit {\it et al} \cite{SGT}. 
The extension to the dynamics was performed by Goldschmidt and Schaub (GS)
\cite{GS}.
Since they presented many details of their  calculations, we shall not repeat
them here. Rather we only outline the approach and provide appendices with 
detailed explanations which complement these given by  GS. 

 The following renormalization constants are defined through the relations between
the bare and the renormalized  couplings:
\begin{equation}
D_{0}=Z_{D}D,\quad 
g_{0}=Z_{g}g
,\label{eq:rgs1}
\end{equation}
\begin{equation}
m_{0}^{2}\phi ^2 =m^2 \phi _{R}^2 ,\quad 
\gamma_{0}^{2}\phi ^{2}=\gamma^2 \phi _{R}^2,\quad
\phi^2 =Z_{\phi }\phi _{R}^2,\quad
\tilde {\phi }^2 =(\tilde {Z}_{\tilde {\phi }}) \tilde {\phi
}_{R}^2,\quad
\label{eq:rgs2}                                    
\end{equation} 
\begin{equation}
\mu _{0}=(Z_{\tilde {\phi }}Z_{\phi } ^{-1})^{-\frac {1}{2}}\mu 
.\label{eq:rgs3}
\end{equation}      
The  subscript $R$ labels the renormalized field variable,
and $0$ the bare variable or the coupling constant.
The terms without subscripts are renormalized constants.                                      
For convenience, we define $\tilde {Z}_{\tilde {\phi}}=[Z_{\tilde {\phi}}]^2$. 

The renormalization of $\mu $ depends on  $Z_{\phi}$ and ${Z}_{\tilde{\phi}}$,
 and no additional  $Z$
factor for renormalization is required.
This is due to  the FDT  which implies:
\begin{equation}
-\theta(t) \frac{d}{dt}\langle \phi(x,t)\phi(0,0)\rangle=\mu
\langle 
\phi(x,t) \tilde {\phi}(0,0)\rangle  . 
\label{eq:fdtt}
\end{equation}
Here $\theta(t) $  equals  to $1$ as $t >0$,  and $0$ as $t <0$.
 Eq.~(\ref{eq:rgs3}) is obtained  by substituting
Eq.~(\ref{eq:rgs2}) into Eq.~(\ref{eq:fdtt}). 
 As we will show later, $D\mu$ will  not suffer any 
renormalization.
Therefore, it is not necessary to calculate $Z_{D}$ in the harmonic
model, which obeys the fluctuation
dissipation  theorem (FDT).

As outlined  by GS the model has a very important 
symmetry $\phi(\vec x ,t) \ra \phi (\vec x , t)           
+ f(x)$, where $f(x)$ is an arbitrary  spatial function constant in                
time.                   
As a result  $\phi (\vec {x},t)$ cannot   be renormalized  
and   $Z_{\phi}=1$ \cite{GS} to all orders in $g$.

We also mention that the lattice effects and the quenched disorder in the substrate 
violate the Galilean symmetry, which
provides another   Ward identity $z+\alpha=2$ \cite{KPZ} 
for the system without these  effects.

The calculations of the $Z$ factors are exemplified  in the appendices
in which the explicit calculations of some of them  are given.

\underline{Recursion Relations}:

Once the Z-factors are known to the leading order in $g$, the
recursion relations are obtained via the so-called $\beta $-functions \cite{Zinn,amit,janssen}:

\begin{equation}   
\beta_{\mu }=\kappa (\frac{\partial \mu}{\partial \kappa})_b  
 =\mu \kappa (\frac {\partial \ln Z_{\tilde{\phi}
}}{\partial \kappa })_{b}=(\frac {g\gamma^2
\sqrt{c}}{D\mu })\mu ,
\label{eq:mu}
\end{equation}    
\begin{equation}
\beta_{D}=\kappa (\frac{\partial D}{\partial  \kappa})_b =-D\kappa (\frac {\partial \ln Z_{D}}{\partial \kappa
})_{b}=(-\frac {\gamma^2 g \sqrt{c}
}{D\mu })D,
\label{eq:dd}
\end{equation}     

\begin{equation}
\beta _g =\kappa (\frac {\partial g }{\partial \kappa})_b
=-g\kappa (\frac {\partial \ln Z_{g}}{\partial \kappa})_b
=2\de g +\frac {2\pi g^2}{(D\mu)^2},
\label{eq:gg}
\end{equation}
\bee
\beta _{\bar{\nu}}=\kappa (\frac{\partial \bar{\nu}}{\partial \kappa })_b
 = -\frac{\pi\gamma ^2}{4 (D\mu)^2}g^2,
\ene
where   subscript $b$   means that all bare parameters are
fixed when one performs the  differentiations \cite{Zinn,amit,janssen} and
 $\kappa$ is a mass scale.                     
The renormalization of the couplings
may also be related to the same $\beta $ functions.
Their flow under a scale change by a factor $b=\exp (l)$
is given by minus the related $\beta $ function, in addition
to the naive dependence which originates in the rescaling of $x\ra bx$ 
 $k\ra b^{-1} k$, and $t\ra b^z t$.  
The recursion relations  so obtained are as follows:
\begin{eqnarray}
\frac {d \nu}{dl}&=&0 , \\
\frac {d \tilde {\bar {\nu}}}{dl}&=&\frac {\pi \gamma ^2}{4\nu
(\tilde {D}\tilde {\mu})}\tilde {g}^2 , \\
\frac {d\tilde {D}}{dl}&=&[2-z +\frac {\tilde {g} \gamma ^2 \sqrt
{c}}{\tilde {D}\tilde {\mu}\nu}]\tilde {D}, \\
\frac {d\tilde {\mu}}{dl}&=-&[2-z +\frac {\tilde {g} \gamma ^2
\sqrt
{c}}{\tilde {D}\tilde {\mu}\nu}]\tilde {\mu}, 
\label{eq:rrmu}\\
\frac {d\tilde {g} }{dl} &=& [2-\frac {\gamma ^2 \tilde {D}\tilde
{\mu}}{2\pi
\nu}] \tilde {g}-\frac {2\pi}{(\tilde {D}\tilde {\mu})^2} \tilde
{g}^2 . \label{eq:rrg}
\end{eqnarray}                                                   
In the next section we discusse  the asymptotic scaling behaviors 
implied by these flow equations.  
                                                                 
\section{Discussion : }
\label{sec5d}

Recalling  that the temperature of the system is $T=\tilde{D}
\tilde{\mu}$, we 
find that $\frac{\partial T}{\partial l}=0 $.
Hence the temperature is not renormalized. The behavior
of the system is governed by the renormalization of the coupling
$g$. The  flow of $\tilde{g}$ depends crucially on  the temperature.
Let us define  $T_{sr}=\frac{\nu b^2}{\pi}$.
The recursion relation for $\tilde{g}$ takes  the form:
\begin{equation}
\frac{\partial \tilde{g}}{\partial l}= 2(1-\frac{T}{T_{sr}})
\tilde{g} -\frac{2\pi }{T^2}\tilde{g} ^2.
\end{equation}

Therefore for $T>T_{sr}$ $\tilde{g}$ flows to zero, while for $T<T_{sr}$
$\tilde{g}$ flows to a fixed point of order $-\delta=1-\frac{T}{T_{sr}}$
with a continuous line of fixed-points [See Fig. 2].

We now analyze the dynamics in each phase separately.

\subsection {The high temperature phases: $T>T_{sr}$}
Since $\tilde{g}\rightarrow 0$ in this phase the equilibrium  properties
are the same as in the high temperature rough phase if a surface
on a  smooth substrate: $C(L)\sim \frac{T}{\nu} \ln (L)$.

Way above $T_{sr}$ the mobility of the surface is finite. However,
as $T_{sr}$ is  approached the mobility becomes  smaller and eventually
vanishes at $T=T_{sr}$.

Integrating the recursion-relation we find that:
\begin{equation}
\tilde{\mu } \sim \tilde{\mu} _0 (\frac{T}{T_{sr}}-1)^{2\sqrt{c}}
\end{equation}
with  $2\sqrt{c}=1.78$.

The dynamic exponent remains $z=2$ throughout this phase, although
the asymptotic scaling behavior is reached only on scales $L>L_g$
where $L_g\sim g_0 ^{\frac{-1}{2\delta}}$ is  the scale on which $\tilde{g}$
decays to zero. The scale $L_g$  diverges as $T\ra T_{sr} $ since
$\ln  (L_g)\sim (\frac{T}{T_{sr}}-1)^{-1} $.

 As 
long as one sits at a temperature $T >T_{sr}$, the decay
 of $\tilde{g}$ under the flow will not alter the asymptotic scaling behavior for $L>L_{g}$, 
except
that the amplitude  in the correlation function depends on 
the bare value of $\tilde{g}$.

In the  high temperature phase, the flow of $\tilde{\mu} (l)$
can  be  calculated in terms of $\tilde{g}(l)$:
\begin{equation}
\tilde {\mu}(l)=\tilde{\mu (0)} e^{-\beta \int _{0}^{l} \tilde{g}(l')dl'},
\end{equation}
where $\beta =\frac {\gamma ^2 \sqrt{c}}{\tilde{D}\tilde{\mu}\nu}$, and 
\begin{equation}
\tilde {g}(l)= \frac {-\de  \tilde {g}(0) e^{-\de  l}}{-\de -
\alpha \tilde{g}(0)+\alpha \tilde{g}(0) e^{-\de l}}
\end{equation}
with   $\alpha =\frac {2\pi}{(\tilde {D}\tilde {\mu})^2}$.

Now we can obtain the macroscopic  mobility, $\tilde{\mu}\sim
\tilde {\mu} (l\ra
\infty)$, in the linear response regime ($F\ra 0$) for the high
temperature phase.
\begin{equation}
\tilde {\mu}(l=\infty)=\tilde {\mu}(l=0) (\frac {|\de|}{\alpha
\tilde {g}(0)+|\de|})^{\beta /\alpha} \sim |\de|^{\beta /\alpha}
,
\end{equation}
where $\frac {\beta}{\alpha}=2\sqrt{c} \sim 1.78$.

As $T\ra T_{sr} ^+$, $\mu _M$  vanishes continuously as
 demonstrated above.
In the low-temperature phase $\tilde{g}(l)$  flows to a finite value. Its
fixed-point location  changes  with the temperature. More explicitly,
the set of fixed points of different temperatures forms a fixed line
in the plane of $\tilde{g}$ and $T$, in which  $\tilde{g}^*(T)\sim T_{sr} -T $, to
first order.
In this phase the scaling equilibrium properties and the
dynamics,
as well as the transport properties, are drastically modified.
Most of the forthcoming discussions are  devoted to this new, 
super-rough phase.

\subsection{The Low Temperature ($T<T_{sr}$) Super-rough Phase}
\label{sublts}

~~~~The regime of temperatures below the transition  provides the
most                                           
exciting new physics. The theoretical predictions are:
(i) The correlations will change from  $C(L)\sim \ln L$
to $C(L)\sim (\ln L)^2$. Hence the surface is even rougher than
that in
the rough phase at $T > T_{sr}$. This behavior was dubbed by Toner
and Divincenzo as super rough. \cite{TD}.
They found it first in  surface of crystals with bulk
disorder.
(ii) The dynamic critical exponent $z$  now displays a temperature
dependence. It increases continuously from its value above
$T_{sr}$
$z=2$, and $z-2$ is to  first order  linear in $T_{sr} -T$.
(iii) The linear response macroscopic mobility vanishes. 
The response becomes   non linear \cite{NG}, at least close to the 
transition, such that the average velocity $v$ scales with
the external force $\tilde{F}$ as $v\sim \tilde{F}^{\eta +1}$, where $\eta $ 
is also a temperature dependent exponent (related by scaling to the
 dynamic exponent $z$).

In the following   we  concentrate on  each of these physical
manifestations separately:

\underline {1. Super-rough  Equilibrium   Correlations}

For $T_{sr} > T $, $g$ approaches a line of fixed points $\tilde{g}^*
= (T_{sr}-T)/\pi$ (see Fig. 2).  
As was shown by Toner and DiVincenzo \cite{TD}, the  
correlation function of the 2-d surface  acquires a second   logarithmic
factor. Since they used  the static replica formalism for a model with
bulk disorder, we explain here how this behavior is obtained within the
 dynamics formualtion, for the case of disorder in the substrate.
 
The Fourier transformation of  $C(L, 0)$ 
2-point vertex function $\Gamma _{0,2} (q,t=0)$ is  defined as before:
\begin{equation}
\Gamma_{0,2} (q, t=0)=<<h(q)h(-q)>>_{disorder}.
\end{equation}  
                         
For systems with vanishing $g$, 
$\Gamma _{0,2}(q)= \frac {1}{\nu q^2}$. Once the term $\bar{\nu}
\int\int \int dx dt dt^{'} \nabla \tilde{h}(\vec x,t)$
$\nabla \tilde{h}(\vec x, t^{'})$
has to be accounted for, the height-height correlation 
function are  obtained from the quadratic part of the Hamiltonian 
which contains this term. 
One  obtains the equal time dynamic correlation,
 (namely the static correlation function):
\begin{equation}
\Gamma _{0,2}(q, t=0) = \frac {1}{\nu q^2}+ \frac {\bar {\nu} q^2 }
{(\nu q^2)^2} .
\label{eq:ccc}
\end{equation}
The first term is the equal time correlation originating
from the time dependent part of the correlation. The second term 
arise from the so called time-persistent part, which will be  
explained in Appendix A.
This  can be viewed as an EA type
order parameter as mentioned  in the previous section. 

The second term is proportional to $\frac {1}{q^2}$ as the first.
 However, because $\bar {\nu} (l)$ increases with $l$ (while
$\nu$ remains unchanged with $l$), it carries another scale
dependence. The height-height correlation function, under
as scale transformation by the factor $b=e^{l}$, changes as:
\begin{equation}
<h(\vec q) h(-\vec q) > = \Gamma _{0,2} (\vec q; \nu , \bar {\nu}(0),
g(0))=e^{2l}\Gamma _{0,2} [e^{l}\vec q ; \nu ,\bar {\nu} , g(l)]  .
\label{eq:vex}
\end{equation} 

The first term in Eq.~(\ref{eq:vex}) comes from the naive dimension of
$<h(\vec q) (-\vec q)>$. 
Using Eq.~(\ref{eq:ccc}) with the renormalized values one obtains:

\begin{equation}
<h(\vec q) (-\vec q)> = e^{2l}\frac {1}{\nu e^{2l}q^2} 
[1+\frac {\bar {\nu}(l)}{\nu}] = \frac {1}{\nu q^2}
[1+\frac {\bar {\nu} (l)}{\nu }].
\end{equation}
 
Apart from a finite part, $\bar {\nu } (l)$
increases as $\sim (g^*)^2 l$.
By choosing $l$ such that $e^{l}q =\Omega =a ^{-1}$, (where
$\Omega$, the momentum cutoff,  can be chosen  as  the inverse  lattice
spacing), the vertex function, $\Gamma _{0,2} (q, t=0)$, is found as:

\begin{equation}
\Gamma _{0,2} (q, t=0) =\frac {1}{\nu q^2}[A-B \frac {(g^*) ^2}{\nu}
\ln (qa)],
\end{equation}
where $A=1, B=\frac {\pi \gamma ^2}{4\nu ^2} $.
Consequently the static correlation function is:
\begin{equation}
C(L)= A^{'} \ln (\frac {L}{a}) + B^{'} \ln ^2 (\frac {L}{a}).
\end{equation} 

Hence, the behavior found by Toner and DiVincenzo   \cite{TD} for the bulk disorder is
also reproduced  in the system under consideration, where only the
substrate is disordered. Experimentally, it might be difficult to
distinguish
 $\ln (\frac {L}{a})$ from  $\ln ^2 (\frac {L}{a})$. However,
the
dynamical behavior in both phases are apparently different and 
their difference may be detected  by the experimental
observations.

The behavior found above persists in the regime $L << \tau ^{1/z}$.

For $L>>\tau ^{1/z}$ simple scaling implies a dependence of $(\ln \tau)^2$
on $\tau$. The derivation of the intermediate behavior of $C(L, \tau)$ is
beyond the scope of this paper (it requires the knowledge of $\Gamma_{0,2}
(\vec q, \omega)$.
However, simple physical considerations hint very strongly that its behavior
 is of the form $C(L,\tau)\sim \{\ln L f[\tau / L^z]\}^2$.

\noindent \underline{ 2. The Dynamic  Exponent }

To calculate the value of the dynamic exponent $z$ in the low
temperature phase  one should look at the
recursion relations of $\tilde{D}$ and $\tilde{\mu}$. 
To locate the fixed point one can require
that  both of $\frac {d\tilde{D}}{dl}$ and $\frac {d\tilde{\mu}}{dl}$ to 
be equal to zero.
Then $z$  is obtained as 
$z= 2+ 4\sqrt {c}|\delta|$,
where the fixed point value of $g$, $g^ {*} = \frac {-\delta }{\pi}$,
was inserted. As far as
we know, this is  the first example in which
the present dynamic exponent $z$  
varies with  temperature continuously (besides the random 
anisotropy XY model, which is described by the same theory
studied by GB). 
 The physical implications  of the increasing $z$ can be understood from
the fact that the relaxation time to reach the equilibrium state
is  longer as the temperature is lowered below $T_{sr}$.
This is to be expected in the phase where the disorder is
relevant, and the surface turns  super-rough as explained in the
previous
section. Since the surface stretches itself to find the configurations
with the lower free energy, it will leave these locations slowly. 
The slower dynamics implies also a   graduated increase in the
averaged free energy $F(L)$ barriers, associated with a scale $L$,
due to the disorder.
                         
\noindent \underline {3. The Non-Linear Response}

In the last section, addressing the high temperature rough phase,
we have found that the linear response mobility vanishes as 
$(T-T_{sr})^{1.78}$, when $T$ is reduced to $T_{sr}$. Below
$T_{sr}$
the linear response mobility $\tilde{\mu }_M$ vanishes. 

As we show in the following the response becomes non-linear.
Again the physical origin of this behavior takes roots
in the preferred configurations of the surface which 
are local minima of its free energy.
Applying a small force $F$ will not move the surface in 
a uniform velocity.
Actually a somewhat similar situation occurs in the 
smooth phase of a surface growing upon a flat substrate,
where the mobility jump from a finite value to zero at the 
roughening temperature. 

If the pinning is even stronger the surface grows by
activation  of higher islands.
If the substrate is disordered,
the preferred and inhomogeneous locations of the surface are
enough to slow the motion and then to cause the linear response
mobility  to vanish.
However, it still allows for a uniform motion
with average velocity which vanishes as $\tilde{F}^{1+\eta }$ $(\eta > 0)$
when $\tilde{F}\ra 0$. Hence, the velocity vanishes faster than $\tilde{F}$.
The force $\tilde{F}$ is a relevant field which increases as 
$\tilde{F}(L)\sim \tilde{F}_0 L^2$ with  the length scale. Even for small $\tilde{F}_0$ 
there is a scale 
$L^* \sim a F_0 ^{-1/2}$ for which the scaled force is of order 
1. Namely it is not a negligible quantity. On the scale,
$L>L^*$ the behavior is not 
"critical". The large force moves
the
surface with a uniform velocity. The ratio between the force
and the velocity is determined by the mobility $\mu$
at the boundary $L^ *$ between the 
 "critical"
($L< L ^*$) and the "non-critical "($L> L ^*$) regimes.
Within the scaling picture
$L^*$ can serve as a "cutoff length ".  
 Hence it is the mobility $\mu (L^ *)$
for a piece of the interface with linear extent $L ^*$ that
is determining the mobility of the whole surface on scales 
$L> L ^*$. Note that $\tilde{\mu} (L)$ is not exactly $\tilde{\mu} (L^*)$
but the ratio between them is  finite  since no 
" critical scaling " holds for $L> L ^*$. The scaling of
$\tilde{\mu} (L) $ for $L<L^*$ may be derived from its definition
as:
\begin{equation}
\frac {\partial h}{\partial t} = \mu F  .                
\end{equation}
 
Under rescaling $L\ra L/b$ we already know that $t\ra t/b ^z$
and $F\ra F b^2$ while $h $ does not suffer any renormalization.
To make both sides scale similarly we must have $\mu \ra \mu b^{2-z}$
and defining $\mu \sim \mu _0 b^{-2\eta }$ we obtain: 

\begin{equation}
\eta =(z-2)/2 =-2\sqrt {c}\delta = 1.78 |\delta |
\end{equation}
That implies that $\mu (L)$ on a scale $L< L ^*$ is
made smaller by $(\frac {L}{a})^{-2\eta}$
with respect to its bare value. In particular we find:
\begin{equation}
\mu (L ^*) = \mu _0 (\frac {L ^*}{a})^{-2\eta }
=\mu _0 F_0 ^{\eta }
=\mu _0 F_0 ^{z/2 -1},
\end{equation}
where we have used relation between $L^ *$ and  $F_0$.

So we have identified the dependence of $\mu _M$
on $F _0$ from which we obtain  the averaged velocity:
\begin{equation}
v\sim \mu _0 F_0 ^{\eta } F_0 = \mu _0 F_0 ^{1+ \eta}
=\mu _0 F_0 ^{z/2}.
\end{equation}

Thus as temperature is lowered the velocity (for the same tiny 
force $F _0$) becomes smaller. How far below $T_{sr}$
these relations hold?
The scaling picture is based on a local equation of motion of
$h$.
Therefore, it implicitly assumes the existence of single solution
for the equation of motion in the limit of vanishing uniform 
force. Just below $T_{sr}$ this is a valid assumption
since
even if more than one minima exists the scale associated with the
difference between the  minimizing 
configurations (which diverges as $T \ra T_{sr}$) is larger than the
scale we
discussed here. At lower temperature this may  no longer hold. 
The existence of multiple minima might  be felt on the
relevant scale (i.e. $L ^*$ ). In the regime where many minima
are
relevant the dynamics will be   activated.
In other words, the slower processes  are going to be related 
to the height of the barriers between these minima.
Some works \cite{toner,nat} have been devoted to estimating these barriers
and drawing the  conclusions based on " activated  dynamics".
It is not clear, however, how reliable these heuristic
estimates are (e.g. the barriers are identified with the
fluctuation in the minima of the free energy.)
Unbounded barriers  between configuration
unrelated by symmetry will also lead to broken ergodity which can be reflected
in "replica symmetry breaking". Such a possibility  was found recently within
a variational approach \cite{korsh}.

\subsection{ T at and just below $T_{sr}$}

The discussion we presented so far for $T<T_{sr}$ applies on length 
scales for  which $\tilde{g}$ is already close to its fixed-point value
$\tilde{g}^*$. If the temperature  is very close to $T_{sr}$ this scales
became very large and it will be necessary to account for the crossover
regime. The recursion relations can be integrated as before and
yield for the scale dependent mobility:
\begin{equation}
\tilde{\mu} (l)= \tilde{\mu}(0) \{1+ \frac{4\pi }{T^2}\tilde{g}(0)  
\frac{1}{2|\delta|}[
(\frac{L}{a})^{2|\delta|}-1]\} ^{-2\sqrt{c}}.
\end{equation}

>From this  expression we see that if $(\frac{L}{a})^{2|\delta|} >> 1$
the results of the previous section apply.
If the force $\tilde{F}$ is not  small enough for the associated scale $L^*$ to
satisfy this relation  the condition  between  the velocity and 
the applied force  changes to:
\begin{equation}
v/F \sim \{1+ \frac{4\pi \tilde{g}(0)} {T^2} \frac{1}{2\delta} (
\tilde{F}^{-|\delta|}-1)
\}^{-1.78}
\end{equation}

At $T=T_{sr} (\delta =0)$ this expression yields:
\begin{equation}
\frac{v}{F}= (1+ \frac{2\pi g(0) }{T^2}\ln  (F) )^{-1.78}
\end{equation}
Namely $v/F $ vanishes as $F\ra 0$ due to logarithmic  corrections.
They originate from the effect of  $g$  which is marginally irrelevant and 
decays to zero so slowly that it still causes $v/F$ to vanish.

\section{Summary}
\label{sec5s}

To summarize the main conclusions of our investigation on the effect
of disorder in the substrate on the surface dynamics:
There exists a phase transition between a high temperature, rough,
phase, and a low temperature, super-rough, phase.

In the rough phase correlations  and response functions have the same
scaling properties as in the pure case. The disorder and the periodic potential
are irrelevant in this phase. As the transition temperature $T_{sr}$ is
approached from above the macroscopic mobility vanishes continuously
as $(T-T_{sr})^{1.78}$ (contrary to the flat substrate in which
it has a finite jump).

The properties of the low temperature phase are unique. The height-height
correlations are 
$C(L,\tau)\sim[\ln L]^2 $ as $ L<<\tau  ^{1/z}$, and 
$C(L,\tau)\sim[\ln \tau]^2 $ as $ L>> \tau ^ {1/z}$ 
with $z=2+4\sqrt{c}(1-\frac{T}{T_{sr}})$  to first order in $(1-T/T_{sr})$.
Namely  the dynamic exponent increase continuously from
its Gaussian value of $2$ as the temperature is lowered below the transition.

The linear mobility has a scale dependence which causes it to
vanish on  large scale. We have shown that this scale dependence
results in a non-linear relation between the applied force and the average
velocity:
$v\sim F^{1+\eta}$, where $\eta =1.78(1-\frac{T}{T_{sr}})$
is also a temperature-dependent exponent.

All these  results also 
apply to the 2D vortex-glass system in  a film of type-II superconductors.

We discussed these implication elsewhere \cite{ST}. The most important one is the non-linear
relation  between the voltage $V$ and the current $I$:
$V\sim I^{1+\eta}$ with $\eta$ given above.

Our RG calculations give the same static behavior obtained by
the replica approach presenting  the unbroken symmetry  between replica.
 
Other  works have shown  the symmetry to be broken within a non-perturbative variational gaussian
approach which is equivalent to the $N\ra \infty$ limit. It is
an unsettled  matter  whether the symmetry is indeed broken for $N=1$.

Preliminary  numerical results for a 2D vortex-glass \cite{Hwaa} and a random growth
model\cite{CS}   show a transition in the dynamic properties at a temperature
within 10 $\%$  of the analytic RG result.  Looking at the static 
correlations below $T_{sr}$, however, the possibility of the replica
symmetry breaking may not   be excluded.  However, It is  too
early to draw firm conclusions
from those preliminary results. It is to be expected, therefore, that
more analytical and numerical  works would be necessary to reconcile the different 
results and to reach a complete  understanding of this exciting and
challenging problem.

\section*{Acknoweldgments}

We are thankful to  M. Kardar, G. Grinstein, B. Schmittmann, T.
Giamarchi, and
especially to D. Huse for most useful discussions. We are also
indebted to  Y. Goldschmidt for bringing to our  attention Ref. \cite{GS}.
We are also thankful for  useful discussion 
D. Cule and T. Hwa regarding their numerical results.

Acknowledgment is also made to the donors of The Petroleum Research Fund, 
administered by the ACS, for support of this research.

\vskip 15pt
\appendix
\section { Response  and correlation functions }
\label{a51}

Two kinds of propagators, the response function and the
correlation function,  arise from the free (Gaussian)  portion in
the effective action in Eq.~(\ref{eq:dgen}) (consisting of
a quadratic form  in  the field  $\phi$ and  the auxiliary field
$\tilde {\phi}$).
One can calculate the free response 
and  correlation functions directly from the free part of the
action $S_0$ in Eq.~(\ref{eq:s02}).
In the momentum and frequency representation, they are:
\begin{equation}
\langle \phi (\vec {q},\omega )\tilde {\phi}(-\vec
{q},-\omega)\rangle =\frac {1}{\mu (q^2+m^2)+i\omega }, 
\label{eq:p1}
\end{equation}
                 
\begin{equation}
\langle \phi (\vec {q},\omega )\phi (-\vec {q},-\omega )\rangle
=\frac {2D\mu ^2}{[\mu (q^2+m^2)]^2+\omega ^2},
\label{eq:p2}
\end{equation}                                 
where $m$ is the mass of the field $\phi$.
 We  have introduced the mass $m$ to regularize the infrared
divergences which appear in the upcoming loop calculations.In the momentum and time representation, they are given by:
\begin{equation}
\langle \phi (\vec {q},t) \tilde {\phi }(-\vec {q},t')\rangle
=\theta (t-t') e^{-\mu (q^2+m^2)(t-t')},
\label{eq:p3}
\end{equation} 

\begin{equation}
\langle \phi (\vec {q},t)\phi (-\vec {q},t')\rangle =\frac {D\mu
}{q^2+m^2}e^{-\mu (q^2+m^2)|t-t'|}.
\label{eq:p4}
\end{equation}

In this manner,  both the  free response  and  free
correlation functions possess  a mass
dependence in their denominators as one can see from Eqs.~($\ref
{eq:p1}$)--Eqs.~($\ref{eq:p4}$).

The short distance cut-off is introduced for regularizing the
ultraviolet  divergence in 2 dimensions. The regularized 
$C_0$ and $R_0$ are:
\begin{equation}
C_{0}(\vec {x},t)=\int \frac {d^2\vec {q}}{(2\pi )^2} e^{i\vec
{q}\cdot \vec {y}-\mu (q^2+m^2)}
\frac {D\mu }{q^2+m^2}|_{y^2=x^2+a^2}
\label{eq:Co}                                            
\end{equation}
\begin{equation}
R_{0}(\vec {x},t)=\frac {\theta (t)}{4\pi \mu t}e^{-\frac
{x^2+a^2}{4\mu t}-m^2 \mu  t}
\label{eq:Re}
\end{equation}
 
Furthermore, the
following equations are very valuable to extract the asymptotic
behavior of  free propagators \cite{GS}:
\begin{equation}
\bar {C}_{0}(x,t)\sim -\frac {1}{4\pi}\ln (m^2_0\la_0|t|)-\frac
{1}{4\pi}C
-\frac {1}{4\pi}\frac {x^2+a^2}{4\la |t|}, \quad \mbox {as} \quad
m^2(x^2+a^2)<<m^2_0\la |t|<<1
\end{equation}
\begin{equation}
\bar{C}_0(x=0,t)= -\frac {1}{4\pi}\ln (2 \sqrt {c} m^{2}_0 \la
|t|)+O(\frac {a^2}{\la_0 |t|}) ,
\end{equation}
where $C_0(\vec x,t)=D\mu\bar{C}_0(\vec x ,t)$,  and $C$ is the Euler's constant
$\approx 0.5772$.
In the limit $m^2_0\la |t|<<m^2(x^2+a^2)<<1$, the equal time
correlation
behaves as:
\begin{equation}
\bar{C}_0(x,t=0)=-\frac {1}{4\pi}\ln (cm^2_0(x^2+a^2))+O(x^2),
\end{equation}

The zero order fluctuation-dissipation theorem (FDT)  relates the
response function  to the
correlation function as follows:
\begin{equation}
-\frac {1}{\mu ^2 D}\theta (t) \frac {d}{dt}C_{0}(\vec
{x},t)=R_{0}(\vec {x},t).
\label{eq:fdt}
\end{equation}    
1.Basic diagrams

  In Fig. 3 the wavy line represents the auxiliary field $\tilde {\phi }$;
  the straight line represents  the field $\phi $.
  The dash line is set to separate two different time
coordinates. The dot points represent
the abbreviation of the other  $\phi$ lines.

2. Free propagators:

 Correlation function  $C_{0}(\vec {x},t)$
and response function $R_{0}(\vec {x},t)$ are shown in Fig. 4.

In general,   the equal time correlation  function,
$ C(\vec x, t=0)=<\phi(\vec x,t)\phi(0,t)>$,
is identical  to the static
correlation function $<\phi(x)\phi(0)>$ averaged by the
Boltzmann weight.
As a special case for the  connection, one can  refer  to 
Eq.~(\ref{eq:p4}). 
On the other hand, the response
function
is related to the reaction of the system to the external
probe, say $P(\vec x,t)$.  The perturbed Hamiltonian will result in
${\cal H}-\int d^d x P(x,t)\phi(x,t)$,
and the additional  term in the Hamiltonian is tantamount to adding $P(x,t) \mu_0
\tilde {\phi} (x,t)$  in the MSR action. In other words,
the conjugate probe  $P(x,t)\mu_0$  will couple to
the response
field $\tilde{\phi}(x,t)$. Intuitively, the linear response function can be
defined as the  ratio of the strength of the reactive field to 
that of the probe:
\begin {equation}
R_{phy}(\vec x-\vec x',t-t')=\frac{\delta \langle
\phi(x,t)\rangle}{\delta P(x',t')}
=\mu_0\frac {\delta \langle \phi(x,t)\rangle }{ \delta \tilde J
(x',t')}
=\mu_0\langle \phi(x,t)\tilde {\phi}(x',t')\rangle .
\label{eq:res}
\end{equation}
where $\mu_0 P(\vec x,t)=\tilde J (x,t)$
(note that $R_{phy}$ is different from  $R_0$ by a factor $\mu_0$ 
and the latter is used in the calculation here). 
This  arises from the various representations of $\tilde{\phi}$,
which is associated with the unphysical degrees of freedom in
multiplying the Langevin equation by an arbitrary constant.                                        

Consistently, the response function will 
manifest  causality. There is no response before the external
probe is applied,
i.e.,            
$R(x,t)=0$ for $t<0$. 

The relation between the static and the dynamic  properties are through the  
correlation functions. In the random systems one distinguishes
between equal-time correlations which correspond to
the static correlations:
\bee
<\phi(\vec x,t)\phi(\vec x^{'}, t)>=[<\phi (\vec x)\phi(\vec x^{'})>]
_{AV}
\ene  
where $[..]_{AV}$ stands for the average over disorder.
In the replica calculation it is given by 
the diagonal term $<\phi _{\alpha} (\vec x) \phi _{\alpha} (\vec x^{'})>
$.
 Time persistent part \cite{Domi} 
of the correlation which correspond to the 
 Edwards-Anderson (EA) type of static  correlation  \cite{sg}:  
\begin{equation}                       
[<\phi (x,t)\phi (x^{'},t^{'}>_{t-t^{'}\ra \infty}]_{AV}=Q_{EA}(x-x^{'}) ,
\end{equation}  
 
In the   replica language, if there is no
replica-symmetry breaking, it is given by the non-diagonal part: 
$\lim _{n\ra 0}[<\phi^{\alpha}(x)\phi^{\beta}(x^{'} )>_{\alpha \neq
\beta}]_{AV}$.

\section { Perturbative Expansions Using Feynman Diagrams }
\label{a52}

To facilitate the RG calculation, one can introduce 
1PI(1-particle-irreducible)
vertex  generating functional $\Gamma [M]$. 
In analogy to the static case, one can define
the vertex functions through $\Gamma$.
For example, the vertex function $\Gamma _{1,1} $
is define by $\frac{\delta ^2 \Gamma [M]}{\delta M_1 \delta M_2}$
\cite{janssen}.
 Here  
the calculation is focused on expanding the bare parameters.
On the other hand, one also can expand the theory in terms of
renormalized parameters. 
Up to 2nd-order perturbation, the 1PI vertex generating
functional are
expanded as \cite{Neu}:                                   
\begin{equation}
\Gamma [M]=\frac {1}{2}(M,G_{0}M) +P[M],
\label{eq:ex1}
\end{equation}
\begin{eqnarray}
P[M]&=&\langle  V[\phi +M]\rangle _{0}
 - \frac {1}{2}\{\langle
V^{2}[\phi  + M]\rangle _{0}-\langle V[\phi +
M]^{2}>\rangle _{0}\}-\{\langle \frac {\delta V[\phi +M]}{\delta \phi }
\rangle _{0}   
,\nonumber \\
&&G_{0}\langle \frac {\delta V[\phi +
M]}{\delta \phi }\rangle _{0} \} +O(V^3), 
\label{eq:ex2}
\end{eqnarray}
where $G_0 ^{-1}$  is symbolizing  the $2\times 2$ free
propagators matrix, 
composed by $R_0$ and $C_0$, and the bracket, $<>_0$, stands for
the average taken  with respect to the free action.     
The  second term in the right hand side of the equation  is the sum of the
connected  diagrams up to the order of $V ^2$, and the third term is
serving to remove all connected diagrams in the second term
which are not 1PIs.

To proceed with the   calculations systemically, it is convenient to
introduce the vertex  functions here. The bare and
renormalized vertex functions can be related by factors of
 $Z_{\phi }$, $Z_{\tilde
{\phi}}$.                       
For instance,
\begin{equation}                                        
\Gamma_{N,L}^{R}(q,\omega; \la _R,m_R,\kappa)=(Z_{\phi })^{\frac     
{L}{2}}(\tilde {Z}_{\tilde {\phi }})^{\frac {N}{2}}\Gamma _{
N,L} (q,\omega;\la_0,m_0,a),
\label{eq:rb}                                 
\end{equation} 
where $\la _R$ and $\la _0$ label  renormalized parameters 
($g_R,\mu,\cdots$) and bare
parameters ($g_0, \mu _0,\cdots$), respectively. 
 $q$ and $\omega$ are the external  momentum and  frequency, respectively.
In the corresponding vertex function, $a$ is a short-distance
cutoff, and $\kappa$ is a mass scale.
Here $\Gamma_{
N,L}$ stands for the vertex function with $L$ external $\phi$ lines
and $N$ external $\tilde {\phi}$ lines. The factors, $Z_{\phi}$ and 
$Z_{\tilde {\phi}}$, are set to remove the divergent parts of
the vertex function $\Gamma$.

\section {The Calculation of  $\tilde {Z}_{\tilde{\phi}}$ } 
\label{subcp}
~~~~Prior to the calculation, we make some remarks about the notations
of the parameters. Since  $m_R$ and $\g _R$ satisfy Eq.~(\ref{eq:rgs1})
and Eq.~(\ref{eq:rgs2}) with $Z_{\phi}$ being $1$, we have
$m_0=m$ and $\gamma _0 = \gamma$. Thus there should be no confusion
if we use them interchangeably.                                                  
Before entering the calculations of the $Z$ factors, one should  make
reference to the relations
between  the various vertex functions. 
For the calculation of the $Z_{\tilde {\phi
}}$ factor, one should focus on  the  vertex  function \cite{GS}, 
$\Gamma _{1,1}$. Inferred from Eq.~(\ref{eq:rb}),
their  renormalized and bared counterparts
are related by:
\begin{equation}
\Gamma ^{R}_{1,1}(q\omega; \mu,\nu,g)= \tilde
{\tilde {Z}}_{\tilde {\phi }}
\Gamma _{1,1} (q\omega; \mu _0 , m_0, g_0)       
=\tilde {Z}_{\phi} \Gamma _{1,1}(q\omega; \mu
Z_{\tilde {\phi }}^{-1}
,m, Z_g g).
\end{equation}
The contribution 
from the action of the first order in $g$ to $Z_{\tilde {\phi}}$ 
can be calculated by
considering:
\begin{equation}
\frac {\delta ^2}{\delta \tilde {M}(\vec {x},t)\delta M(\vec
{0},0)}\langle V_{g}[\phi +M, \tilde{\phi}+\tilde{M}]\rangle 
,\label{eq:gr1}                          
\end{equation} 
where
\begin{equation}
V_{g}[\tilde {\phi },\phi ]=-\frac {\mu ^2 \gamma^2 g}{2a^2}\int
\int d^2x dt dt' \tilde {\phi }(\vec
{x},t)\tilde {\phi }(\vec {x},t')\cos [\gamma (\phi (\vec
{x},t)-\phi (\vec {x},t'))]
.\label{eq:gr2}
\end{equation}

 Fig. 5.
  shows one representative of the diagrams corresponding
to $\Gamma_{1,1}$ in Eq.~(\ref{eq:vex2}). The dots inside the circle
stand for the contractions of  $\phi(\vec x,t)$s and $\phi (\vec x ,0)$s, 
which results in $e^{\g ^2 C_0 (0,t)}$ in Eq.~(\ref{eq:vex2}).
The  dots near the left "ears" represent the contractions of 
$\phi(\vec x ,t)$s and $\phi(\vec x,t)$s, which result in  
$e^{-\g ^2 C_0 (0,0)}$ in Eq.~(\ref{eq:vex2}). The roles of the dots 
near  the right "ears" are similar.

In the frequency representation, the integral coming from
the contractions of inner lines can be written as: 
\begin{eqnarray}
 &&\g ^2  \{ \int _{-\infty}^{\infty}e^{i\omega t}dt
R(0,t) e^{-\g ^2
C_0(0,0)+\g ^2 C_0(0,t)} \nonumber \\
&&\quad -(i\omega) \frac {1}{\mu ^2 D}\int _{0}^{\infty} dt e^{i\omega
t}e^{-\gamma ^2 C_0(0,0)} \}
\label{eq:vex2}\\
&&= e^{-\g ^2 C_0(0,0)}\int _{0}^{\infty}dt e^{i\omega t}dt
\frac {1}{\mu ^2 D}\frac {d}{dt} \g ^2 C_0(0,t) [e^{\g ^2
C_0(0,t)}-1
]\nonumber  \\
&&= e^{-\g ^2 C_0(0,0)}i\omega \int _{0}^{\infty}dt e^{i\omega t}
[e^{\g ^2 C_0(0,t)}-1]
(\frac {1}{\mu ^2 D})
\end{eqnarray}
Combing with the prefactors and using $g=g_0(cm^2 a^2)^{\delta} $, 
we have
\begin{eqnarray}
&& (-i\omega) \g ^2 c m^2 \frac {\mu_0 ^2}{\mu_0 ^2 D_0}g
\int _{a^2/\la}^{1/m^2 \la}\frac {dt}{2\sqrt {cm^2 }\mu_0 t}.
\nonumber \\
&&
\end{eqnarray}
Finally, $\tilde {Z}_{\tilde {\phi}}$ is found as:
\begin{equation}
\tilde {Z_{\tilde {\phi}}}=1+ \frac {g\g ^2 \sqrt {c} }{2 * (\mu
D)}
\ln (\kappa ^2 a^2),
\end{equation}
where $\kappa=cm^2$ is the scale, at which one can 
impose the prescription 
for the vertex functions  corresponding to the renormalized
parameters \cite{Neu}.

\section{The Calculation of $Z_{D}$} 
\label{subcd}
~~~~To calculate the factor $Z_D$, we consider the vertex
function
$\Gamma _{2,0}$.
The bare vertex function $\Gamma _{2,0}$
is related to $\Gamma _{2,0}^{R}$ through
$Z_{\tilde {\phi}}$. They can easily be seen from Eq.~(\ref{eq:rb}):        
\begin{equation}
\Gamma_{2,0}^{R}(\vec {q},\omega ;\mu ,g
)=\tilde
{Z}_{\phi } ^2\Gamma_{2,0}^{0}(\vec {q},\omega ;\tilde {Z}_{\tilde
{\phi}}\mu
,Z_{g}g ).
\label{eq:ftr}
\end{equation}

The associated diagram is illustrated in Fig. 6.
After taking the contraction of the inner lines, we are left
with:
\begin{equation}
-\frac {g\mu _0 ^2 \g _0 ^2}{2a^2}\int \int  \tilde {M
}(x,t^{'}) \tilde {M}(x,t),
\label{eq:cont}
\end{equation}
times $2<\cos \{\g[\phi (x,t)-\phi (x,t^{'})]\}>_0$, where
the contents inside the bracket are averaged with respect
to the free action.
The latter term  contributes to the renormalization of two-point 
vertex depicted in  Fig. 6.
The calculation of this term can be performed with ease  as shown
below.
In the time representation, it reads:
\begin{equation}
-2\frac {g\mu _0 ^2 \g _0 ^2}{2a^2}e^{-\g ^2 C_0(0,0)}\int
_{-\infty}^{\infty}dt
e^{\g ^2C_0 (0,t)} e^{i\omega t}.
\end{equation}
In the limit $\omega \ra 0$, it reduces to:
\begin{equation}
-2\frac {g\mu _0 ^2 \g _0 ^2}{2a^2}e^{-\g ^2 C_0(0,0)}\int
_{-\infty}^{\infty}dt
e^{\g ^2 C_0(0,t)}. 
\end{equation}
With the aid of the formula given in Appendix A, the divergent part
of
this term can be obtained by proceeding in the same manner as in
the  previous section:
\begin{equation}
-2\frac {g\mu _0 ^2 \g _0 ^2}{2a^2}e^{-\g ^2 C_0(0,0) }\frac
{-2}{2\sqrt {c}m^2 \mu}
\ln (cm^2a^2) + \mbox {finite terms}.
\end{equation}
Then, 
\begin{equation}
-2D\mu ^2 = (\tilde {Z}_{\tilde {\phi } })^2 \{ -2D_0 \mu _0 ^2
+\g
^2 \sqrt c g \mu _0
\ln (cm^2a^2)\}. 
\label{eq:eud}
\end{equation}
Finally we inherit $Z_D=1+  \frac {\g ^2 \sqrt
{c}g}{2D\mu}\ln (cm^2 a^2)$.  There should be no confusion
if we put either the  bare  or  the renormalized  values of $D$ and $\mu$ in
the denominator in the above equation. Up to this order, there
will be no difference between these two  possibilities.

APPENDIX E: THE EXPRESSION OF $Z_{g}$  AND THE RENORMALIZATION OF  $\bar{\nu}$
\label{subcg}

 It is a  tedious calculation to find $Z_g$ by
considering  the vertex function $\Gamma _{2,0
}(\vec {q},\omega )$.  Since  quenched disorder is
present  
in the system, one may instead consider $\Gamma_{2,0}(\vec
{q},t_{1};-\vec q,t_2) $,
in the limit $|t_{1}-t_{2}|\rightarrow \infty$.

To obtain the scaling equation of $\bar{\nu}$,   one  should consider  
the $\vec q$ dependent part of  $\Gamma_{2,0}(\vec
{q},t_{1};-\vec q,t_2) $ with $|t_{1}-t_{2}|\rightarrow \infty$. 
The contractions of fields which connect the points $t_1$ and $t_2$, 
make no  contributions.   
The $\vec q$ dependent part of $\Gamma _{2,0}$ up to order $g^2$  involves
7 diagrams, which are  classified into 3 sets as illustrated in GS \cite {GS}. 
 Those terms can be expressed  as  the first seven terms
in (7.3) of  GS \cite {GS}.
The first set 
(referred to  fig1. (a) in GS)  contains
only one diagram as shown  in Fig 7 (A).  
The second set contains three diagrams, corresponding to   part (B), (C), and (D) in Fig. 7. 
The  third set can 
be obtained  from the second set by  swapping  $ (x_1,t_1) \leftarrow
\rightarrow  (x_2,t_2)$ and  $ (x_1,t^{''})\leftarrow
\rightarrow  (x_2,t^{'})$. 

In the following, 
we only present the calculations which are not explictly  given  
in GS \cite{GS}.   
 In the calculation, we only concentrate  on  the time dependent part, 
neglecting
the prefactors. This  should  not be a cause for  confusion when one  retrieves
them later. 
By changing the variables $\tau ^{'} =t_1 - t^{'}$ and $ \tau ^{''}=
t_2 -t^{''}$  the first  term   is recast into:  
\begin{equation}
\int _{-\infty}^{\infty} d\tau ^{'} \int _{-\infty}^{\infty} d\tau ^{''}
R_0(x, \tau^{'})  R_0(x, \tau ^{''}) 
\sinh \gamma ^2 [ - C_0(x, \tau ^{'})- C_0(x, \tau ^{''})],
\label{eq:g1}
\end{equation}

where $x=x_1-x_2$.

With the identity  of FDT for $R_0$  and $C_0$  and the integration by parts,
it is simplfied into:
\begin{eqnarray} 
\frac{-1}{(\mu ^2 D \gamma ^2)^2}&\cdot&\int ^{\infty}_{0}  \int ^{\infty}_{0}   d \tau ^{'}  d \tau ^{''} 
\{ [\frac{d}{d\tau ^{'}  } \sinh  \gamma ^2 C_0(x, \tau ^{'} ) ]
 [\frac{d}{d\tau ^{''}  } \cosh \gamma ^2 C_0(x, \tau ^{''} ) ] \nonumber \\
&+& [\frac{d}{d\tau ^{'}  } \cosh \gamma ^2 C_0(x, \tau ^{'} ) ]    
 [\frac{d}{d\tau ^{''}  } \sinh  \gamma ^2 C_0(x, \tau ^{''} ) ] 
\} \nonumber \\ 
&=&  \frac{-2}{(\mu ^2 D \gamma ^2)^2}  \sinh ( \gamma^2 C_0(x,0))
[\cosh  (\gamma^2 C_0(x,0))-1].
\label{eq:gg1}
\end{eqnarray}

By changing the variables $\tau ^{'} =t_1 - t^{'}$ and $ \tau ^{''}=
t^{'}-t^{''}$ in the second term (see part (B) in Fig.7 ) and  
$\tau ^{'} = t^{''}-t^{'}$ and $ \tau ^{''}=
t_2 -t^{''}$ in 5th term,  they become:
\begin{equation}
-\int _{-\infty}^{\infty} d\tau ^{'} \int _{-\infty}^{\infty} d\tau ^{''}
R_0(x, \tau^{'})  R_0(x, \tau ^{''}) \exp [ \gamma ^2 C_0(0, \tau ^{'}
+ \tau ^{''} )]  \sinh  \gamma ^2 [ C_0(x, \tau ^{'})- C_0(x, \tau ^{''})].
\label{eq:g2}
\end{equation}

By changing the variables $\tau ^{'} =t_1 - t^{'}$ and $ \tau ^{''}=
t^{'}-t^{''}$ in the third term (see part (C) in Fig. 7) and  
$\tau ^{'} =t_2 - t^{''}$ and $ \tau ^{''}=
t^{''}-t^{'}$ in the 6th term, they become:
\begin{equation}
-\int _{-\infty}^{\infty} d\tau ^{'} \int _{-\infty}^{\infty} d\tau ^{''}
R_0(x, \tau^{'})  R_0(0,\tau ^{'} + \tau ^{''}) \exp [ \gamma ^2 C_0(0, \tau ^{'}
+ \tau ^{''} )]  \cosh \gamma ^2 [ C_0(x, \tau ^{'})- C_0(x, \tau ^{''})].
\label{eq:g3}
\end{equation}

By changing the variables $\tau ^{'} =t^{''}- t^{'}$ and $ \tau ^{''}=
t^{'}-t_1$ in the 4th term (see part (D) in Fig. 7) and  
$\tau ^{'} =t^{'} - t^{''}$ and $ \tau ^{''}=
t^{''}-t_2$ in the 7th term, they become:
\begin{equation}
\int _{-\infty}^{\infty} d\tau ^{'} \int _{-\infty}^{\infty} d\tau ^{''}
R_0(x, \tau^{'})  R_0(0,-\tau ^{'} - \tau ^{''}) \exp [ \gamma ^2 C_0(0, \tau ^{'}
+ \tau ^{''} )]  \cosh \gamma ^2 [ C_0(x, \tau ^{'})- C_0(x, \tau ^{''})].
\label{eq:g4}
\end{equation}

With the FDT identity, Eq.~(\ref{eq:g2})   can be 
futher simplfied into:
\begin{eqnarray}
-\frac{1}{(\mu ^2 D \gamma ^2)^2}\int _{0}^{\infty}  d \tau ^{'} \int _{0}^{\infty}  d \tau ^{''}
&&\exp [\gamma ^2 C_0(0, \tau ^{'} + \tau ^{''} )]
[\frac{d}{d \tau ^{'} } \cosh  \gamma ^2 C_0(x, \tau ^{'} ) 
\frac{d}{d \tau ^{''} }  \sinh  \gamma ^2 C_0(x, \tau ^{''} ) \nonumber  \\
&-& \frac{d}{d \tau ^{''}} \cosh \gamma ^2 C_0(x, \tau ^{''} )
\frac{d}{d \tau ^{'}  }  \sinh  \gamma ^2 C_0(x,  \tau ^{'} )].
\label{eq:g5}
\end{eqnarray}

By using the FDT and the integration by parts, one can reduce Eq.~(\ref{eq:g3})  into:

\begin{eqnarray}
\frac{1}{(\mu ^2 D \gamma ^2)^2}\int _{0}^{\infty}  d \tau ^{'} \int _{-\tau^{'}}^{\infty}  d \tau ^{''}
&& \exp [\gamma ^2 C_0(0, \tau ^{'} + \tau ^{''} )]
[\frac{d}{d \tau ^{'} } \cosh  \gamma ^2 C_0(x, \tau ^{'} ) 
\frac{d}{d \tau ^{''} }  \sinh  \gamma ^2 C_0(x, \tau ^{''} )  \nonumber  \\
&-& \frac{d}{d \tau ^{''}} \cosh \gamma ^2 C_0(x, \tau ^{''} ) 
\frac{d}{d \tau ^{'}}  \sinh  \gamma ^2 C_0(x,  \tau ^{'} )]  \nonumber  \\
&+& \mbox { the boundary term }.
\label{eq:g6}
\end{eqnarray} 
  
Simliarly, Eq.~(\ref{eq:g4})   is recast  into: 
\begin{eqnarray}
\frac{1}{(\mu ^2 D \gamma ^2)^2}\int _{0}^{\infty}  d \tau ^{'} \int _{-\infty}^{- \tau ^{'} }  d \tau ^{''}
&&\exp [\gamma ^2 C_0(0, \tau ^{'} + \tau ^{''} )]
[\frac{d}{d \tau ^{'} } \cosh  \gamma ^2 C_0(x, \tau ^{'} ) 
\frac{d}{d \tau ^{''} }  \sinh  \gamma ^2 C_0(x, \tau ^{''} ) \nonumber  \\
&-& \frac{d}{d \tau ^{''}} \cosh \gamma ^2 C_0(x, \tau ^{''} )
\frac{d}{d \tau ^{'}}  \sinh  \gamma ^2 C_0(x,  \tau ^{'} )] \nonumber  \\
 &+& \mbox {the  boundary term }.
\label{eq:g8}
\end{eqnarray} 
  
Combing Eq.~(\ref{eq:g5}), Eq.~(\ref{eq:g6}) and Eq.~(\ref{eq:g8}) terms excluding the boudary terms,  
we have:
\begin{eqnarray}
\frac{1}{(\mu ^2 D \gamma ^2)^2}\int _{0}^{\infty}  d \tau ^{'} \int _{-\infty}^{0}  d \tau ^{''}
&\cdot& \exp [\gamma ^2 C_0(0, \tau ^{'} + \tau ^{''} )]
[\frac{d}{d \tau ^{'} } \cosh  \gamma ^2 C_0(x, \tau ^{'} ) 
\frac{d}{d \tau ^{''} }  \sinh  \gamma ^2 C_0(x, \tau ^{''} ) \nonumber  \\
&-& \frac{d}{d \tau ^{''}} \cosh \gamma ^2 C_0(x, \tau ^{''} )
\frac{d}{d \tau ^{'}}  \sinh  \gamma ^2 C_0(x,  \tau ^{'} )]  .
\label{eq:g9}
\end{eqnarray} 
 
The above integral vanishes since the integrand is antisymmetric
with respect to the transformations $ \tau ^{'} \rightarrow 
- \tau ^{''} $ and  $\tau ^{''} \rightarrow  -\tau ^{'}$.

The boundary terms in Eq.~(\ref{eq:g3}) 
and in Eq.~(\ref{eq:g4})  cancell each other also.
For example, the boundary  term in Eq.~(\ref{eq:g3})   reads as:
\begin{eqnarray} 
&&\int _{0}^{\infty}  d \tau ^{'} \frac{d}{d \tau ^{'}  } 
[\sinh \gamma ^2 C_0(x, \tau ^{'} )
] + \int _{0}^{\infty}  d \tau ^{'} \exp [ \gamma ^2 C_0(0,0) ] \nonumber  \\
&&[ \frac{d}{d \tau ^{'} } \sinh \gamma ^2 C_0(x, \tau ^{'} ) 
\cosh \gamma ^2 C_0 (x, - \tau ^{'} )
- \frac{d}{d \tau ^{'} } \cosh \gamma ^2 (x, \tau ^{'} ) 
\sinh \gamma ^2 C_0(x, - \tau ^{'} )] .
\label{eq:g10}
\end{eqnarray}    
 
To sum up,  the only contribution  is from Eq.~(\ref{eq:gg1}).  
  Extracting  the singular contribution 
\cite {GS,amit}
 gives $\bar {\nu}_b = \bar{\nu} + \frac {\gamma ^2g^2}{8\pi(D\mu)^2}
\ln(\kappa ^2 a^2)$.

\newpage

\begin{center}
{\bf  Figures Caption}
\end{center}

\vskip 0.1in
\underline{Fig. 1}

A two-dimensional cut (along a lattice plane perpendicular to the disordered

substrate) of the three-dimensional system.

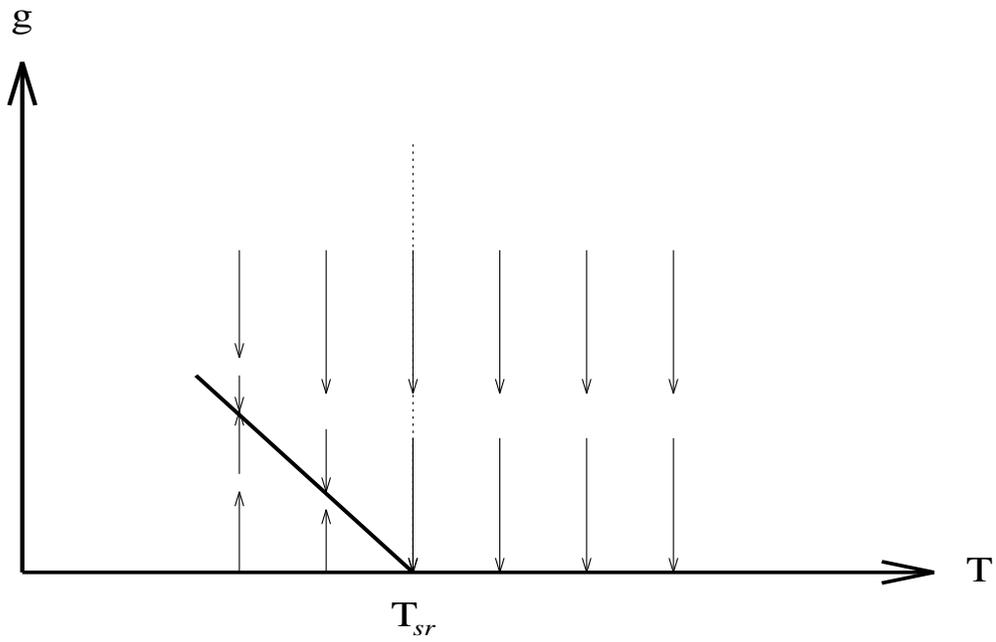
\underline{Fig. 2}

The flow diagram for $g(l)$ with two different fixed-lines for $T$ larger
and smaller than $T_{sr}$.

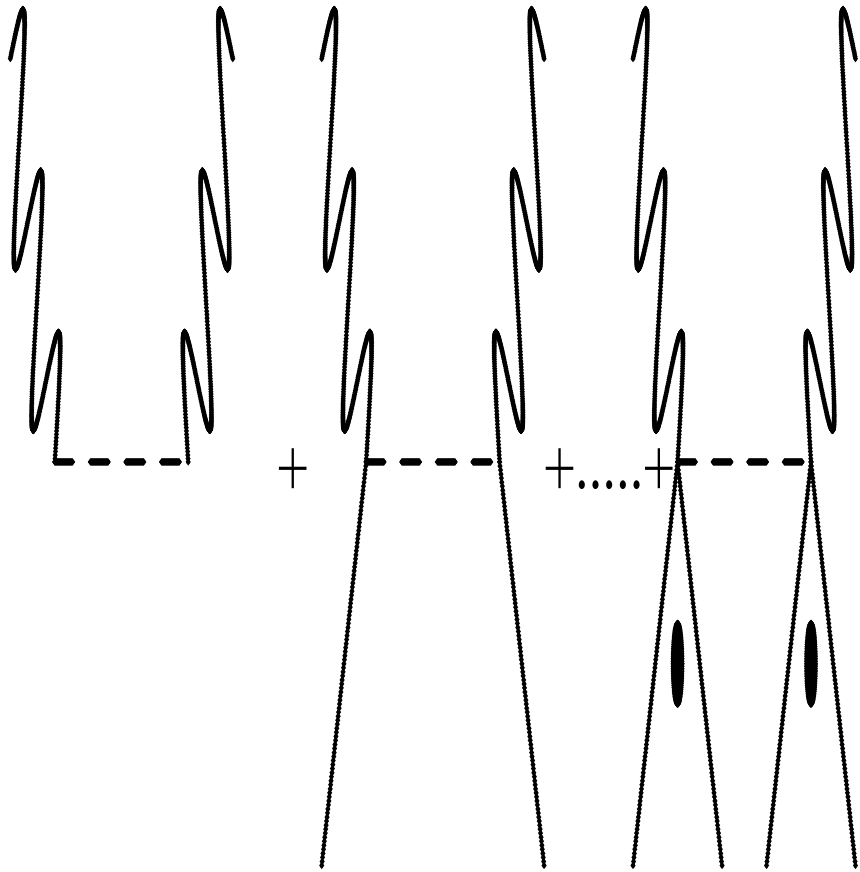
\underline{Fig. 3}

The Feynman diagram representing $\tilde{\phi} (\vec x,t) \tilde{\phi}(\vec x,
t') \cos (\phi(\vec x ,t)- \phi(\vec x, t'))$.

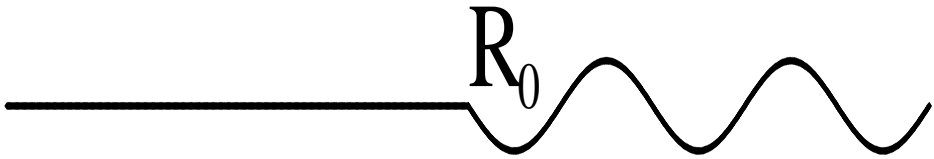
\underline{Fig. 4}

The lines representing the free correlation and response functions.

\underline{Fig. 5}

The Feynman diagram for the vertex function $\Gamma _{1,1}$.

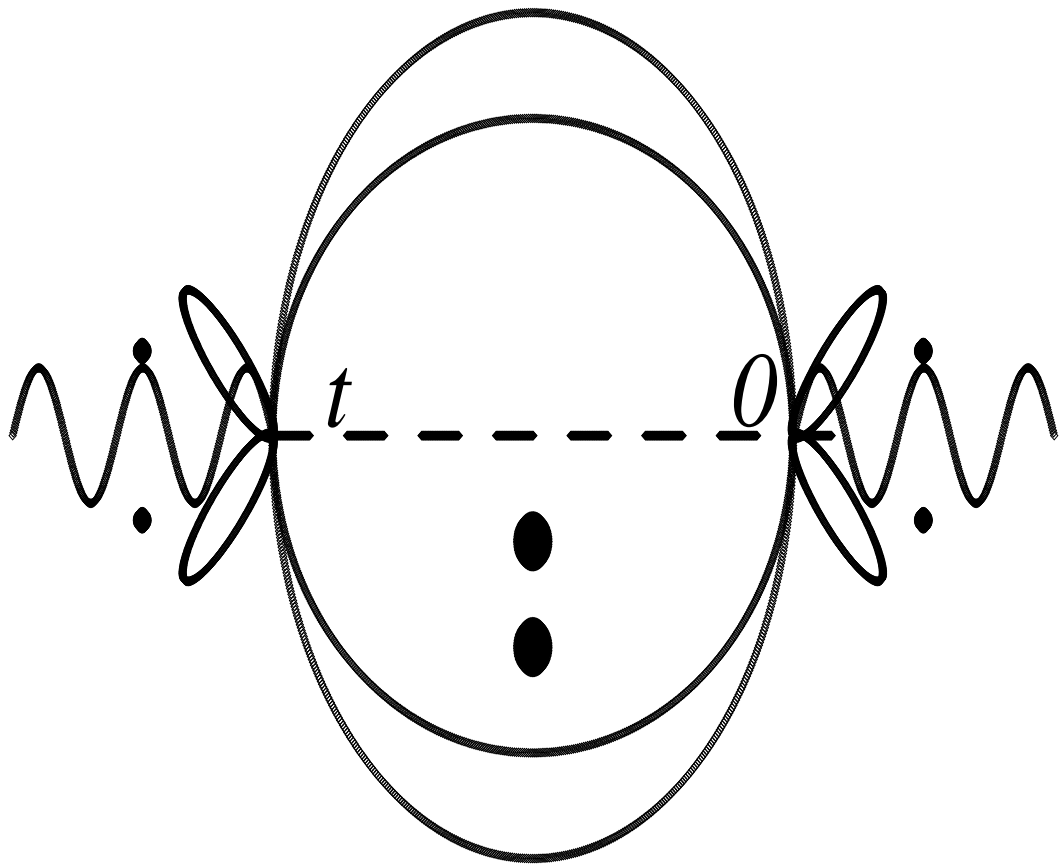
\underline{Fig. 6}

The Feynman diagram for the vertex function $\Gamma _{2,0}$.

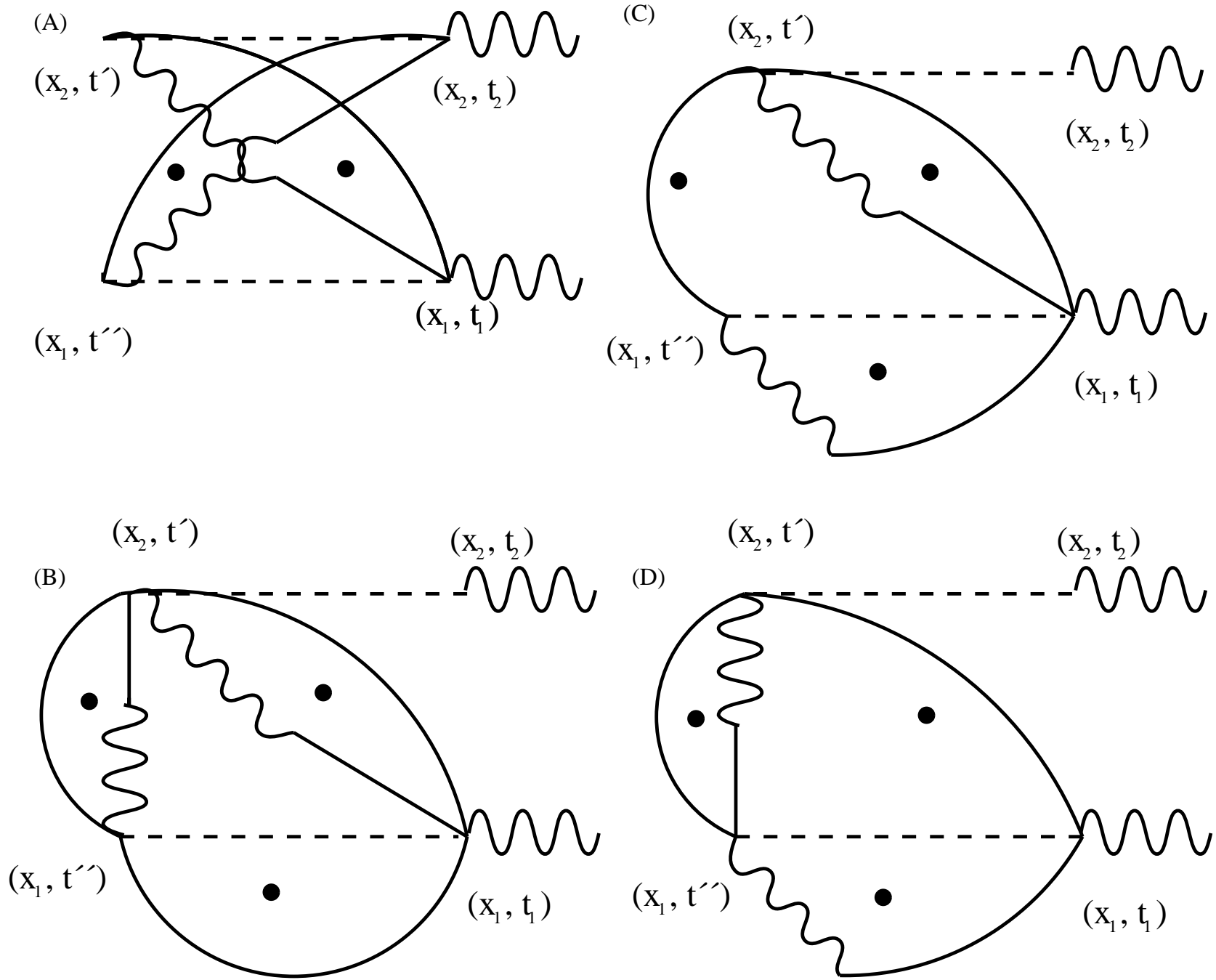
\underline{Fig. 7}

4 Feynman diagrams  contributing  to the renormalization of $\bar{\nu}$.

\begin{thebibliography}{400}
\label{bib}
\bibitem{BN} For a recent review, see H. van Beijeren and I. Nolden, in
 {\it Structure and Dynamics of Surface II}, edited by Schommers and P.von
 Blackenhagen (Springer-Verlag, Berlin, 1987).
\bibitem{GR1} For reviews of recent experimental and theoretical
developments  see, e.g.,  Kinetics of Ordering and Growth  at
Surfaces,  edited  by  M.  Lagally  (Plenum,  New  rork,
    1990); Dynamics of Fractal Surfaces, edited by F. Family
     and T. Vicsek (World Scientific, Singapore, 1991), and
    references therein.
\bibitem{krug} For a review, see e.g.,
 J. Krug and H. Spohn, in {\it Solids far from Equilibrium}, ed.
C. Godreche
(Cambridge University Press, 1991).
\bibitem{CW} S.  T.  Chui  and J.  D.  Weeks, Phys.  Rev.  Lett.
{\bf 40},  733 ( 1978).
 \bibitem{NG} P.  Nozieres  and  F.  Gallet,  J.  Phys.  (Paris)
{\bf 48},  353
   (1987).
\bibitem{TS} Y. C. Tsai and Y. Shapir Phys. Rev. Lett. {\bf 69} 1773 (1992)
\bibitem{MSR} P. C. Martin, E. Siggia, and H. Rose, Phys. Rev. A
{\bf 8}, 423
     (1973).
\bibitem {MM}S. K. Ma and G. F. Mazenko,  Phys. Rev. B {\bf 11}, 4077, (1975).
\bibitem {SMS} S. K. Ma, {\it Modern Theory of Critical Phenomena},
(W. A. Benjamin, New York, 1976);
D. Forster, {\it Hydrodynamic Fluctuation, Broken Symmetry, and Correlation
Functions} (W. A. Benjamin, New York, 1975);
L. E. Reichl, {\it A Modern Course in Statistical Physics}
(University of Texas Press, 1980).
\bibitem{KPZ} M. Kardar, G. Parisi,
and Y. C. Zhang, Phys. Rev. Lett. {\bf 56},
889 (1986); E. Medina, T. Hwa, M. Kardar,
 and Y. C. Zhang, Phys. Rev. {\bf A
39}, 3053 (1989).
\bibitem{FFH89} M. P. A. Fisher, Phys. Rev. Lett., {\bf 62} 1415
(1989).

\bibitem{FFH891} D.~S.~Fisher, M.~P.~A.~Fisher, and D.~A.~Huse, Phys. Rev. B
{\bf 43}, 130 (1990).
\bibitem{nat} T. Nattermann, I. Lyuksyutov, and M.
Schwartz, Europhys. Lett. {\bf 16}, 295 (1991);
T. Nattermann and I. Lyuksyutov, Phys. Rev.
Lett. {\bf 68}, 3366 (1992).
\bibitem{ST} Y. Shapir and Y. C. Tsai , Phys. Rev. Lett. {\bf 71}, 2348
(1993).
\bibitem{EW} S. Edwards and D. R. Wilkinson, Proc. R. Soc. Lond.
{\bf A}  381, 17 (1982).
\bibitem{SGT} D. J. Amit, Y. Y. Goldschmidt, and G. J. Grinstein,
 J. Phys. {\bf A 13}, 585 (1980).
\bibitem{GS} Y. Y. Goldschmidt and B. Schaub,
Nucl. Phys. {\bf B 251}, 77 (1985).
\bibitem{Zinn} J. Zinn-Justin, {\it Quantum Field Theory
 and Critical Phenomena} (Oxford Science Publications, 1989).
\bibitem{amit} D. J. Amit, {\it Field Theory, the Renormalization
Group, and
Critical Phenomena}, (World Scientific, Singapore, 1984).
\bibitem{janssen} H. K. Janssen, Z. Phys. {\bf B 23}, 377 (1976);
B.Bausch, H.K.Jassen, and H.Wanger, Z. Phys. {\bf B 24}, 113 (1976);
C. De Dominicis, and L. Peliti, Phys. Rev. {\bf B 18}, 353 (1978).
\bibitem{TD} J. Toner and  D. P.  DiVincenzo, Phys.  Rev.  B {\bf 41},
    ( 1990).
\bibitem{toner} J. Toner, Phys. Rev. Lett. {\bf 67}, 2537
(1991); J. Toner, {\it ibid} {\bf 68}, 3367 (1992).
\bibitem{korsh} S. E. Korshunov,
Phys. Rev. {\bf B 48}, 3969  (1993);
T. Giamarchi and P. LeDoussal, Phys. Rev. Lett. {\bf 72 }, 1530 (1994).

\bibitem {Domi} C. De Dominicis, Phys. Rev. {\bf B 18}, 4913 (1978).
\bibitem{sg} For a review, see K. Binder and A. P. Young,
Rev. Mod. Phys. {\bf
58}, 801 (1986); and  M. M\'ezard, G. Parisi, and M. A. Virasoro,
{\it Spin Glass Theory and Beyond}, (World Scientific, Singapore,
1987).
\bibitem {Neu} B. Neudecker, Z. Phys. {\bf B 49}, 57 (1982);
{\it ibid} {\bf B 48}, 149 (1982).
(1993).
\bibitem{Hwaa}G. G. Batrouni  and T. Hwa (to be published)
\bibitem{CS} D. Cule  and  Y. Shapir (to be published)
\end{thebibliography}
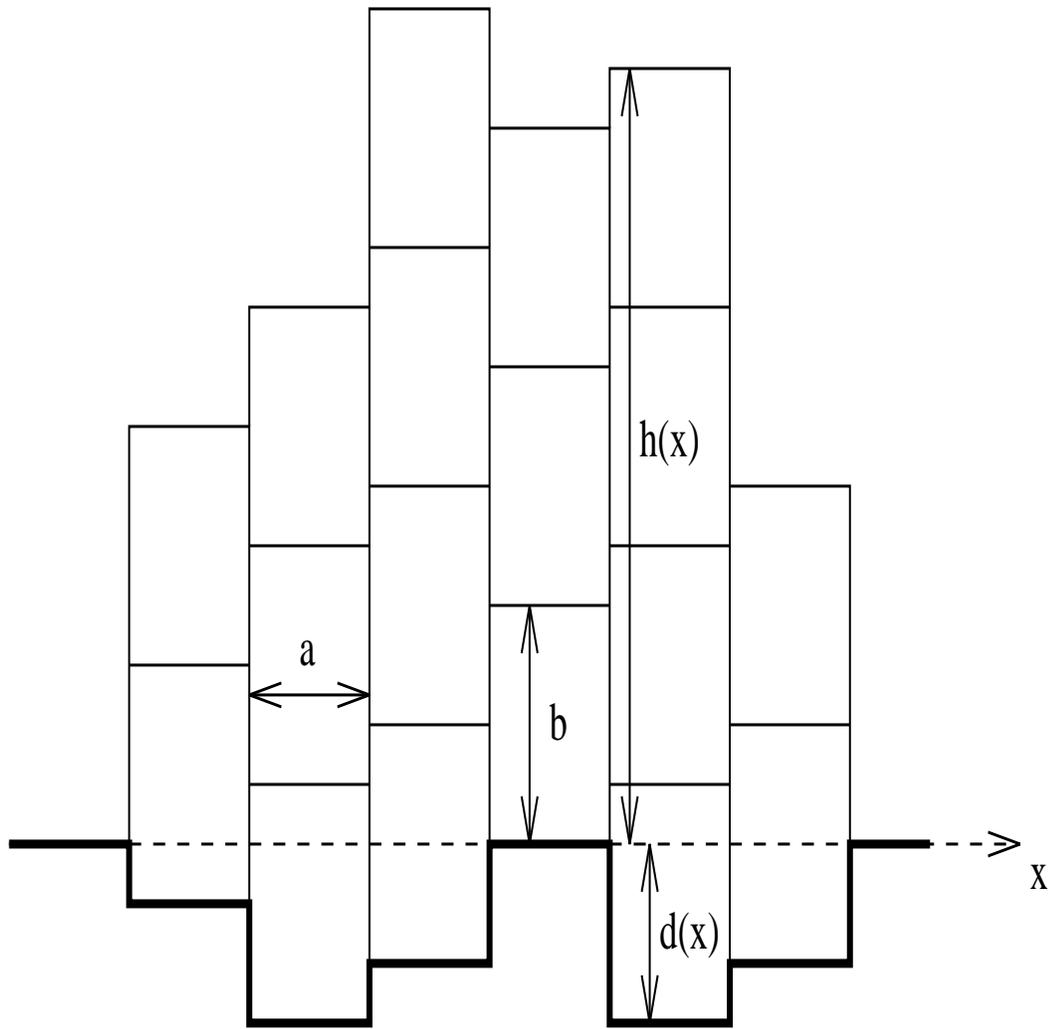
\end{document}